\documentclass[letterpaper]{article} 
\usepackage{aaai2026}  
\nocopyright
\usepackage[T1]{fontenc}   
\usepackage{enumitem}
\usepackage{lmodern} 
\usepackage{multirow, booktabs}
\usepackage{xcolor}
\definecolor{dart}{HTML}{0072B2}   
\definecolor{drift}{HTML}{D55E00}  
\usepackage{times}  
\usepackage{helvet}  
\usepackage{courier}  
\usepackage[hyphens]{url}  
\usepackage{graphicx} 
\usepackage{natbib}  
\usepackage{caption} 
\usepackage{algorithm}
\usepackage{algpseudocode} 

\usepackage{amssymb}
\usepackage{amsfonts} 
\usepackage{amsmath} 

\newcommand{\mbf}{\mathbf}

\newcommand{\mcl}{\mathcal}

\def\R{\mathbb R}
\def\C{\mathbb C}

\newcommand*{\norm}[1]{\left\|#1\right\|}

\newcommand*{\card}[1]{\left|#1\right|}

\newcommand\blfootnote[1]{%
  \begingroup
  \renewcommand\thefootnote{}%
  \footnotetext{#1}%
  \addtocounter{footnote}{-1}%
  \endgroup
}
\usepackage{newfloat}
\usepackage{listings}
\DeclareCaptionStyle{ruled}{labelfont=normalfont,labelsep=colon,strut=off} 
\floatstyle{ruled}
\newfloat{listing}{tb}{lst}{}
\floatname{listing}{Listing}
\newcommand{\AB}[1]{\textcolor{black}{#1}}
\newcommand{\LV}[1]{\textcolor{black}{#1}}

\title{Physics-Guided Diffusion Priors for Multi-Slice Reconstruction in Scientific Imaging}
\author{Laurentius Valdy$^{1}$,
Richard D.~Paul$^{1,2}$,
Alessio Quercia$^{1,3}$, 
{Zhuo Cao}$^{1}$, \\Xuan Zhao$^{1}$,
{Hanno Scharr}$^{1}$, 
{Arya Bangun}$^{1}$}
\affiliations{{$^{1}$IAS-8, Forschungszentrum Jülich, Jülich, Germany} \\
{$^{2}$Dept. of Statistics, LMU Munich, Munich, Germany}\\ 
{$^{3}$Dept. of Computer Science, RWTH Aachen University, Aachen, Germany}}

\usepackage{bibentry}

\begin{document}

\maketitle
{\blfootnote{\hspace{-0.6cm}A. Q. and R. D. P. were funded by the Helmholtz School for Data Science in Life, Earth,
and Energy (HDS-LEE)}} 
\begin{abstract}
Accurate multi-slice reconstruction from limited measurement data is crucial to speed up the acquisition process in medical and scientific imaging. However, it remains challenging due to the ill-posed nature of the problem and the high computational and memory demands. We propose a framework that addresses these challenges by integrating partitioned diffusion priors with physics-based constraints. By doing so, we substantially reduce memory usage per GPU while preserving high reconstruction quality, outperforming both physics-only and full multi-slice reconstruction baselines for different modalities, namely Magnetic Resonance Imaging (MRI) and four-dimensional Scanning Transmission Electron Microscopy (4D-STEM). Additionally, we show that the proposed method improves in-distribution accuracy as well as strong generalization to out-of-distribution datasets. 
\end{abstract}

%
\section{Introduction}

Reconstructing multi-slice images has become increasingly important in scientific disciplines, such as medical imaging, plant science, biology, and materials science \cite{midgley20033d, lee2023multislice, schulz20123d, chung2023solving, bangun2025reg, liu2024refining}, mainly to enhance the visualization, analysis, and interpretation of the complex object structures. A persistent challenge in MRI is the inherently slow data acquisition process, which is further compounded by the growing demand for 3D object reconstruction through multi-slice \cite{zbontar2018fastmri, bangun2025reg}. Hence, the main challenge in accelerating MRI is to reconstruct multi-slice images from under-sampled measurements. Similarly, in four-dimensional scanning transmission electron microscopy (4D-STEM) only intensity measurements are recorded in electron detector, thereby creating a demand to retrieve 3D phase information of the object under investigation, i.e., electrostatic potential of materials, through multi-slices model \cite{maiden2012ptychographic,bangun2022inverse,diederichs2024exact}. Most proposed model-based methods for multi-slice reconstruction for both MRI and 4D-STEM, highly \LV{dependent} on the initialization of the algorithm, yielding poor flexibility and often leading to suboptimal reconstructions.

Diffusion models have become powerful tools for generating both 2D and 3D images, achieving notable success in computer vision, data augmentation, and image/video generation \cite{ho2020denoising, ho2022video, sohl2015deep, song2019generative, song2020improved, song2020score}. These models excel at capturing complex data distributions and generating high-fidelity images by iteratively refining \LV{samples} from a random distribution into coherent structures. Despite their success with natural images, applying diffusion models to scientific images is challenging due to their unique characteristics, which arise from specialized modalities such as microscopy, spectroscopy, or medical imaging. 
The incorporation of \LV{multi-slice} reconstruction requirements further amplifies these challenges. In particular, scientific data requires volumetric consistency, robustness to acquisition physics, and interpretability, which conventional diffusion frameworks do not provide directly. Therefore, applying diffusion models directly to scientific imaging raises questions about their suitability and whether they can enhance reconstruction quality while enabling fast image generation. 

We propose a framework that integrates partitioned diffusion priors with physics-based forward models, providing scalable training and physics-consistent inference for multi-slice reconstruction, namely DART: Diffusion-Alternating Multi-slice Reconstruction Technique and DRIFT: Diffusion-Refined Initialization for Multi-slice Reconstruction.  We empirically show that our algorithms are efficient in computation and memory usage (more than $8\times$) \LV{and} robust to out-of-distribution (OOD) data.
{\let\thefootnote\relax\footnote{{\hspace{-0.5cm}Code: \url{https://jugit.fz-juelich.de/ias-8/Distributed_3DDM}}}}
\section{Background}
\begin{figure*}[!htb]
\vspace{-1ex} 
    \centering
\includegraphics[width=1\linewidth]{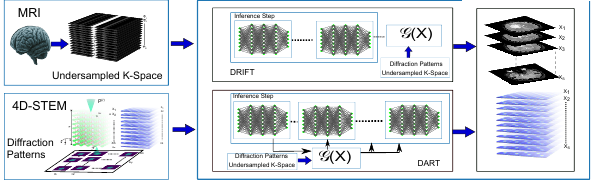}
    \caption{Data acquisition of a brain in MRI (k-space) and 4D-STEM (diffraction patterns) of crystalline materials, as well as proposed methods: DART (alternating update between trained diffusion prior and physics constraints)  and DRIFT (diffusion priors as initialization before applying physical constraints). Physical constraints $\mcl{G}(\emph{X})$ are adapted depending on the modality (MRI or 4D-STEM).}
     \vspace{-2ex} 
    \label{fig:visual}
\end{figure*}
\subsection{Diffusion Models}
Diffusion models use a neural network $\mcl{S}_{\theta}\left(\textbf{\emph X}_t,t\right)$ to learn original data $\textbf{\emph X}_0$, where training data is perturbed at each step $t$ with variance $\beta_t$, $$
\textbf{\emph X}_t = \sqrt{1 - \beta_t} \textbf{\emph X}_{t-1} + \sqrt{\beta}_t \textbf{\emph Z}_{t-1}, \;\; t \in [T],$$
with $\textbf{\emph Z}$ a Gaussian noise. The trained model with optimal parameters $\theta^*$ is used in inference via sampling 
\cite{ho2020denoising,ho2022video}
\begin{equation}
\label{eq:sampling_general}
\textbf{\emph X}_{t-1} = \frac{1}{1- \beta_t}\left(\textbf{\emph X}_t + \beta_t \mcl{S}_\theta^*\left(\textbf{\emph X}_t,t\right)\right) + \sqrt{\beta_t } \textbf{\emph Z}_t.
\end{equation}Intuitively, diffusion models learn to iteratively denoise random noise into structured data samples. This makes diffusion models attractive as generative priors for scientific imaging.
\subsection{Physics of MRI and 4D-STEM}
In many physical modalities, due to accelerated data acquisition or physical limitations of the measurement process, only limited data can be acquired. For example, to accelerate 3D MRI acquisition, undersampled k-space data are collected and can be modeled as $\mbf{\hat{Y}}_s = \mbf{M} \circ \mbf{Y}_s$, where $\mbf{M}$ selects rows/columns from full k-space $\mbf Y_s \in \C^{N \times N}$ for each slice $s \in \{1,2,\hdots S\}$.  

A similar condition appears in multi-slice 4D-STEM, where only intensity measurement can be recorded. In multi-slice 4D-STEM, the exit wave after the first slice of crystalline materials $\mbf{X}_1 \in \C^{N \times N}$is given by
$$\mbf{E}_1^{(r)} = \mbf{X}_1 \circ \mbf{P}^{(r)},$$ where $\mbf{P}^{(r)}$ is the focused electron beam at scan point $r$ \AB{in two-dimensional coordinate axes} with element-wise product $\circ$.  
The exit wave is propagated through vacuum distance between slices using Fresnel operator $\mcl{V}$, $$\mbf{E}_s^{(r)} =  \mbf{X}_s \circ \mcl{V}\left(\mbf{E}^{(r)}_{s-1}\right), s \in \{2,\dots,S\}.$$
At the last slice $S$, the exit wave $\mbf{E}^{(r)}_S$ produces diffraction patterns with Fourier transform $\mcl F$, recorded as
$$\mbf{I}^{(r)} = \card{\mcl{F}\left(\mbf{E}^{(r)}_S\right)}^2 \in \R^{N \times N}.$$ The acquisition process for both undersampled MRI and 4D-STEM are shown in Figure \ref{fig:visual}.
 
The goal in both multi-slice MRI and 4D-STEM is to reconstruct multi-slice object $\emph{X} \in \C^{S \times N \times N}$ given incomplete measurement data, i.e., undersampled k-space in MRI or phase retrieval problem from projection intensities with total scan $R = S_x \times S_y$ in 4D-STEM.  
The optimization problem for 4D-STEM in terms of Frobenius norm is
\begin{equation}
\small
{ 
 \textbf{\emph X$^*$} = \underset{\textbf{\emph X} \in \C^{S \times N \times N}}{\text{arg min}}
\sum_{r = 1}^R \norm{\mbf{I}^{(r)} - \card{ \mcl{F}\left( \mcl{H}\left(\textbf{\emph X},\mbf{P}^{(r)}\right)\right) }^2}_F^2
+ \mcl{R}\left(\textbf{\emph X} \right), 
 }
\label{eq:opt_stem}
\end{equation}
where $\mcl H$ models slice interaction and Fresnel propagation in forward 4D-STEM multi-slice, $\mbf{P}^{(r)}$ is the known focused probe, and $\mcl R$ is a regularizer.  
For MRI, the corresponding optimization is
\begin{equation}
\small
\label{eq:opt_mri}
 \textbf{\emph X$^*$} = \underset{\textbf{\emph X} \in \C^{S \times N \times N}}{\text{arg min}}
\sum_{s = 1}^S \norm{\hat{\mbf{Y}}_s - \mbf{M} \circ \left( \mcl{F} \left( \mbf{X}_s \right)\right) }_F^2
+ \mcl{R}\left(\textbf{\emph X} \right),
\end{equation}
In both settings, incomplete measurements (missing k-space lines in MRI, missing phase information in 4D-STEM) make recovery particularly challenging. Moreover, solutions to these optimization problems depend heavily on initialization, motivating the integration of diffusion priors trained on relevant data distributions.
We denote optimization problems \eqref{eq:opt_stem} \eqref{eq:opt_mri} as $\mcl{G}\left(\textbf{\emph X} \right)$. 
\section{Methods}  
We propose a physics-guided partitioned diffusion inference framework for multi-slice reconstruction \AB{by incorporating trained neural network to generate multi-slice priors. Since the training is straightforward by partitioning the multi-slice data of the MRI object, i.e., brain, and the electron microscopy object, i.e., crystalline materials, we focus on the inference aspect first and defer the training procedure to \LV{the} numerical section}. 

Measurement data (MRI k-space or 4D-STEM diffraction patterns) are incorporated through the respective physical models, ensuring that inference respects instrument constraints. 
\begin{figure*}[ht!]
    \centering    
    \includegraphics[scale=0.9]{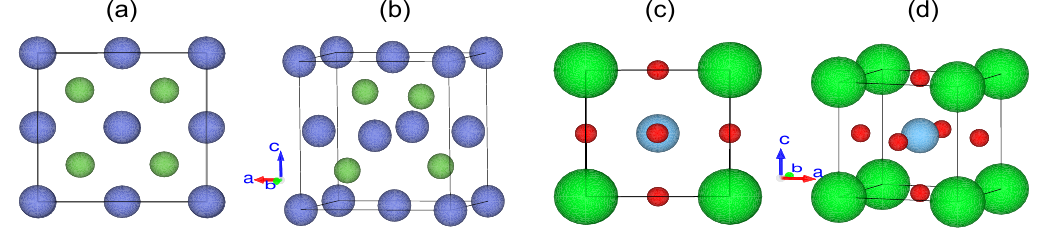}
    \caption{\AB{Example of cubic crystal materials:}
     unit cells of gallium arsenide (GaAs) \AB{with volume dimension $5.6533^3 \AA^3$} (a),(b) projection and 3D visualization, strontium titanate (SrTiO{$_3$}) \AB{with volume dimension   $3.905^3 \AA^3$} (c), (d) projection and 3D visualization. In the multi-slice method, each slice is obtained by calculating the projection of each atomic plane along the z direction.}
      \vspace{-2ex} 
    \label{fig:materials}
\end{figure*}
\begin{algorithm}[!ht]
\caption{Physics-Guided Partitioned Diffusion Inference (\textcolor{drift}{DRIFT} / \textcolor{dart}{DART}) }
\label{algo}
\begin{algorithmic}[1]
\Require $\#$GPUs $G$, model $\mcl{G}$, measurement data $\mbf{\hat Y}_s$ (MRI) or $\mbf{I}^{(r)}$ (4D-STEM)
\State Initialize $\mathbf{X} =
\begin{cases}
  \textcolor{drift}{\mathbf{X} \in \C^{L \times S \times N \times N} \quad (\textsc{DRIFT})}\\
  \textcolor{dart}{\mathbf{X} \in \C^{S \times N \times N} \qquad (\textsc{DART})}
\end{cases}$

\For{$t = T \;\mathbf{to}\; 1$}
  \State Slice size $B_g = \lceil S/G \rceil$
  \ForAll{GPU $g = 0 \dots G-1$ in parallel}
    \State Block $S_g = \min((g+1) B_g, S) - g B_g$
    \State Extract $\mathbf{X}_g \in \C^{S_g \times N \times N}$ or  \State \qquad \quad$\mathbf{X}_g \in \C^{L \times S_g \times N \times N}$
    \State $\mathbf{X}_g \gets \texttt{DDPMSampler}(\mathbf{X}_g, t)$  in \eqref{eq:sampling_general}
    \State Send $\mathbf{X}_g \to$ root
  \EndFor
  \State Gather all $\mathbf{X}_g \to \mathbf{X}$
  \If{mode = DART}
      \State \textcolor{dart}{$\mathbf{X} \gets \mcl{G}(\mbf{\hat Y}_s\,\text{or}\, \mbf{I}^{(r)},\mathbf{X})$} \Comment{apply physics model}
  \EndIf
\EndFor

\If{mode = DRIFT}
    \State \textcolor{drift}{$\mathbf{X} \gets \underset{l \in [L]}{\mathrm{argmax}}\, \texttt{SSIM}(\mbf{\hat Y}_s\,\text{or}\, \mbf{I}^{(r)}, \mathbf{X})$}
    \State \textcolor{drift}{$\mathbf{X} \gets \mcl{G}(\mbf{\hat Y}_s\,\text{or}\, \mbf{I}^{(r)},\mathbf{X})$} \Comment{physics refinement}
\EndIf

\State \Return $\mathbf{X} \in \mathbb{C}^{S \times N \times N}$
\end{algorithmic}
\end{algorithm}
To scale inference across multiple GPUs, $S$ slices are distributed uniformly over $G$ GPUs ($\lceil S/G \rceil$ each), enabling parallel \LV{processing} and efficient memory usage. Our approach includes two complementary variants:  
\begin{itemize}
    \item \textbf{DART} (Diffusion-Alternating Multi-slice Reconstruction): alternates between diffusion prior updates and physics-based constraints at each iteration.  
    \item \textbf{DRIFT} (Diffusion-Refined Initialization for Multi-slice Reconstruction): uses diffusion priors to generate a high-quality initialization, followed by physics-based refinement.  
\end{itemize} The inference procedure is summarized in Algorithm~\ref{algo}, as well as in Figure \ref{fig:visual} where the main difference between DART and DRIFT is visualized. 
In DRIFT model\LV{, the $\texttt{SSIM}$ function is calculated} to measure the similarity of the initialization with the  measurement data. For MRI, the SSIM is evaluated after inverse Fourier transform of undersampled k-space $\mbf{\hat Y}_s$ per slice, i.e., $\mcl{F}^{-1}\left( \mbf{\hat Y}_s \right)$. On the contrary, in 4D-STEM, the bright field image is calculated by integrating or summing the detector image to have two-dimensional data related to the scan dimension before calculating SSIM, i.e., sum over last two axes of dimension $(S_y,S_x, N, N)$ to get $(S_y, S_x)$.
 
The operator $\mcl{G}(\mathbf{X})$ enforces measurement consistency with the underlying physical model. 
For MRI, this corresponds to a \emph{data-consistency step} that replaces the sampled entries in k-space with the acquired measurements:
\begin{equation}
\footnotesize
\mathbf{X}_s^{(k+1)} = \text{Prox}\left(\mathbf{X}_s^{(k)} - \lambda \,\nabla_{\mbf{X_s}}\sum_{s = 1}^S \norm{\hat{\mbf{Y}}_s - \mbf{M} \circ \left( \mcl{F} \left( \mbf{X}_s \right)\right) }_F^2 \right),
\end{equation}
where $\mcl{F}$ denotes the Fourier transform, $\mathbf{M}$ is the sampling mask, and $\lambda$ is a step size. 

For 4D-STEM, $\mcl{G}$ corresponds to one step of gradient descent on the intensity-based loss in Eq.~\eqref{eq:opt_stem}:
\begin{equation}
\fontsize{8.pt}{9pt}\selectfont 
\emph{X}^{(k+1)} = \text{Prox}\left(\emph{X}^{(k)} - \eta \,\nabla_{\emph{X}} 
\sum_{r=1}^R \left\|\mathbf{I}^{(r)} - \big|\mcl{F}(\mcl{H}(\emph{X}^{(k)}, \mbf{P}^{(r)}))\big|^2\right\|_F^2\right),
\end{equation}
where $\eta$ is the step size \AB{\LV{and} the strategy to choose the step size is given for instance in \cite{xu2018accelerated}}. The function $\text{Prox}$ is the proximal function for regularization, for instance, soft thresholding for the sparse structure of the data \cite{parikh2014proximal}. 

In practice, we use a fixed number of physics-update iterations per diffusion step. For instance, in DART approach for every diffusion step we perform one step gradient update. In DRIFT, we perform $100$ steps gradient update after initialization from the diffusion model for complete step $T$. This modular design allows the same framework to generalize across different modalities by substituting the appropriate physical constraint $\mcl{G}$.

\section{Numerical Experiments}
We evaluate our methods on two scientific imaging applications: MRI and 4D-STEM. \AB{Apart from the contribution for inference algorithms, DART and DRIFT, diffusion model is also trained with multi-slice MRI and crystalline materials data.}
\begin{table*}[!ht]
\centering
\caption{Mean and standard deviation of SSIM from BraTS data and cubic crystal data. The \textbf{bold} and \underline{underline} represent the best and second-best results.}
\label{tab:mean_ssim_mri}
\begin{tabular}{cccc}
\hline
Dataset & Mask Types & Methods & SSIM ($\uparrow$) \\ 
\hline
\multirow{5}{*}{$\underset{\text{MRI}}{\text{BRATS}}$}  
& \multirow{5}{*}{$\underset{\substack{2\times, 0.15}}{\text{Uniform}}$} 
& DART & \textbf{0.968} $\pm$ 0.011 \\
& & DRIFT ($L=16$) & \underline{0.938} $\pm$ 0.023 \\
& & Projection-Based \cite{bangun2025reg} & 0.844 $\pm$ 0.027 \\
& & CS MRI \cite{lustig2007sparse} & 0.804 $\pm$ 0.025 \\
& & TV \cite{block2007undersampled} & 0.803 $\pm$ 0.024 \\
\hline
\multirow{5}{*}{$\underset{\text{Crystal Data}}{\text{Cubic}}$} 
& & DART & 0.607 $\pm$ 0.406 \\
& & DRIFT ($L=16$) & \textbf{0.899} $\pm$ 0.091 \\
& $-$ & Sparse Decom \cite{bangun2022inverse} & 0.831 $\pm$ 0.198 \\
& & 3PIE \cite{maiden2012ptychographic} & \underline{0.887} $\pm$ 0.243 \\
& & Torchslice \cite{diederichs2024exact} & 0.823 $\pm$ 0.115 \\
\hline
\end{tabular}
\end{table*}
 \begin{figure*}[htb!]
    \centering 
     \includegraphics[scale=0.75]{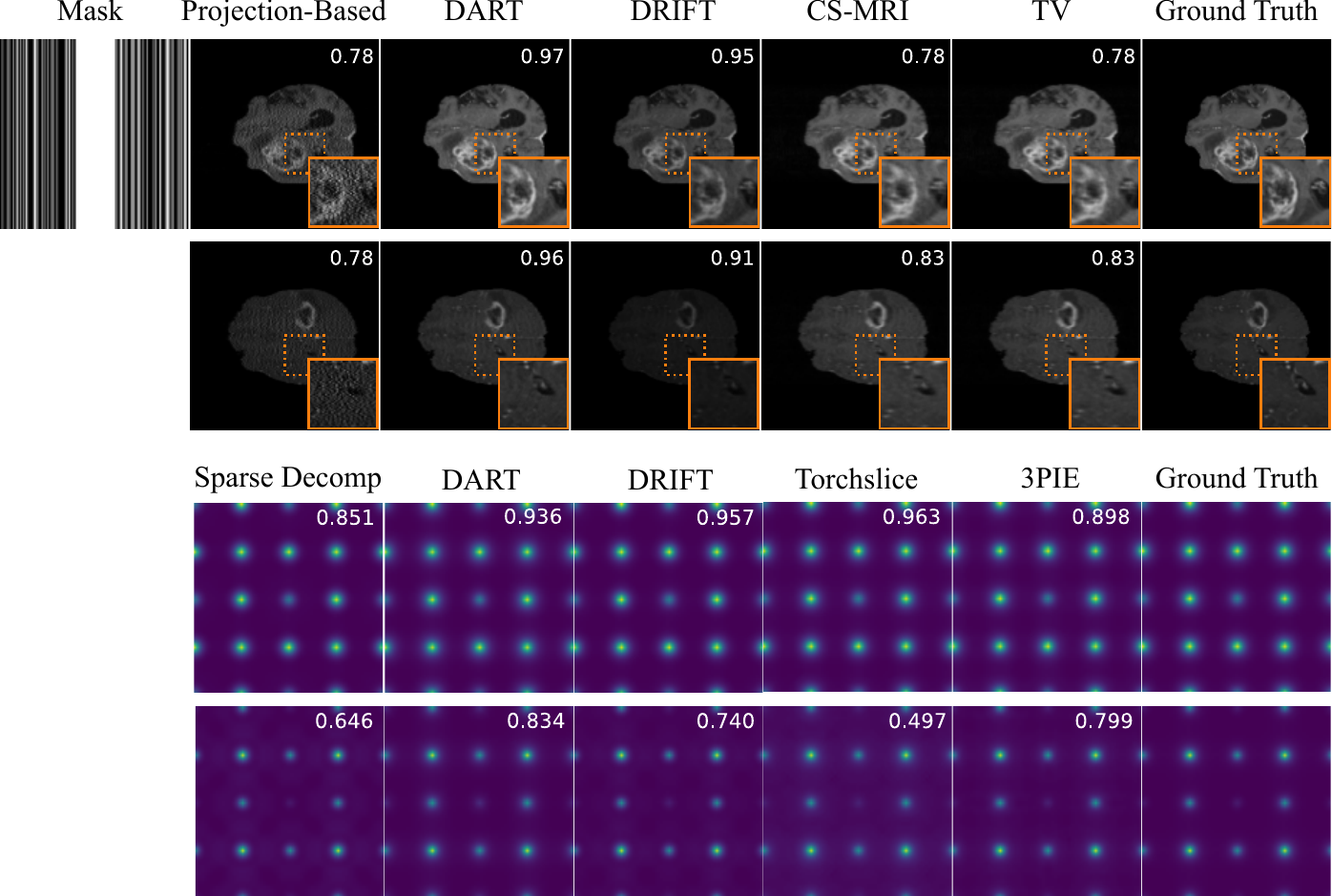}
     \caption{
     Single slice MRI  with zoomed-in region of interest from the volume reconstruction of file BraTS20 Training 338 t1ce (top) and BraTS20 Training 039 t1ce (bottom);  Projection-based \cite{bangun2025reg}; DART; DRIFT; CS MRI \cite{lustig2007sparse}; Total Variation \cite{block2007undersampled}. Phase Projection of crystalline materials CoPt$_3$ (top) and  Tb$_3$InC  (bottom) benchmarking with Sparse Decomposition \cite{bangun2022inverse}; DART; DRIFT; Torchslice \cite{diederichs2024exact}; 3PIE \cite{maiden2012ptychographic}.Top right are visual quality metrics, namely  SSIM.}
    \label{fig:BraTS_Crystal_result}
\end{figure*}
\subsection{\AB{Training Procedures}}
For MRI, we use T2-FLAIR Brain Tumor Segmentation (BraTS) 2020 \cite{bakas2017advancing, bakas2018identifying, menze2014multimodal}   with dimension\LV{s} $155 \times 240 \times 240$. For 4D-STEM, we use a cubic crystal system with lattice structure from the Materials Project \cite{jain2013commentary}. Multi-slice algorithm implemented in \cite{durham2022accurate} is used to generate slices of atomic potentials. The dataset contains pixel dimensions $10 \times 80 \times 80$ per file with a total of $3533$ files. A visualization of the material can be seen in Figure \ref{fig:materials}.
\begin{table*}[ht!]
\centering
\caption{Mean and standard deviation of SSIM from OOD data, namely Roots and Hexagonal Crystal Data. The \textbf{bold} and \underline{underline} represent the best and second-best results.}
\label{tab:mean_ssim_mri_ood}
\begin{tabular}{cccc}
\hline
Dataset & Mask Types & Methods & SSIM ($\uparrow$) \\ 
\hline
\multirow{5}{*}{$\underset{\text{MRI}}{\text{Roots}}$}  
& \multirow{5}{*}{$\underset{\substack{8\times, 0.08}}{\text{Gaussian}}$} 
& DART & \textbf{0.813} $\pm$ 0.130 \\
& & DRIFT ($L=16$) & \underline{0.750} $\pm$ 0.174 \\
& & Projection-Based \cite{bangun2025reg} & 0.611 $\pm$ 0.169 \\
& & CS MRI \cite{lustig2007sparse} & 0.676 $\pm$ 0.151 \\
& & TV \cite{block2007undersampled} & 0.577 $\pm$ 0.163 \\
\hline
\multirow{5}{*}{$\underset{\text{Crystal Data}}{\text{Hexa}}$} 
& & DART & 0.588 $\pm$ 0.355 \\
& & DRIFT ($L=16$) & \textbf{0.981} $\pm$ 0.010 \\
& $-$& Sparse Decom \cite{bangun2022inverse} & \underline{0.977} $\pm$ 0.011 \\
& & 3PIE \cite{maiden2012ptychographic} & 0.953 $\pm$ 0.041 \\
& & Torchslice \cite{diederichs2024exact} & 0.573 $\pm$ 0.025 \\
\hline
\end{tabular}
\end{table*}
\begin{figure*}[!htb]
    \centering
    \includegraphics[scale=0.75]{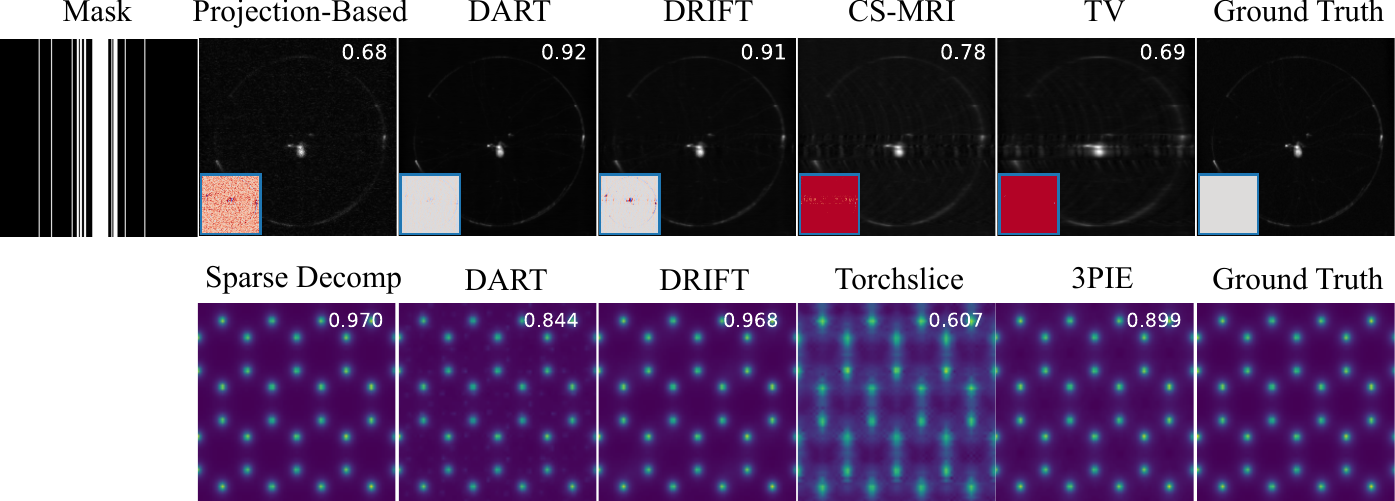}
    \caption{
     Single slice MRI  from the volume reconstruction of file soybean roots;  Projection-based \cite{bangun2025reg}; DART; DRIFT; CS MRI \cite{lustig2007sparse}; Total Variation \cite{block2007undersampled}. Phase Projection of crystalline materials WSe$_2$ benchmarking with Sparse Decomposition \cite{bangun2022inverse}; DART; DRIFT; Torchslice \cite{diederichs2024exact}; 3PIE \cite{maiden2012ptychographic}. Top right are visual quality metrics, namely  SSIM.  }
    \label{fig:ood_results}
\end{figure*}
We use a video diffusion model \cite{ho2022video} that extends 2D U-Nets using convolutions and attention mechanisms that operate separately over spatial and temporal dimensions, to efficiently capture both the appearance and motion present in video data, which is aligned with \LV{both} multi-slice MRI and crystalline materials. \AB{The model is used with base width 64 and channel multipliers (1, 2, 4, 8). Additionally, it is trained with \(700{,}000\) steps using \(\ell_1\) loss, learning rate \(10^{-4}\), batch-level gradient accumulation over 2 steps, and other hyperparameters following \cite{ho2022video}.} To address the computational complexity of training, we adopt a  slice-wise training strategy. The axial slices of MRI and crystal data are partitioned into $G$ groups with respect to the number of GPUs and each assigned to an independent diffusion model. To guarantee the reconstruction of each measurement data for MRI (k-space) and 4D-STEM (diffraction patterns), it should be incorporated in the inference process, as presented in Figure \ref{fig:visual}.
\subsection{Magnetic Resonance Imaging}  We report numerical results in terms of SSIM in Table \ref{tab:mean_ssim_mri} evaluated on 30 random multi-slice data.  Figure \ref{fig:BraTS_Crystal_result} shows the slice reconstruction for various  algorithms. It can be seen that DART achieves higher SSIM values for reconstructing slices and preserves anatomical details close to the ground truth despite the undersampled k-space using a uniform mask. 
Overall, these results demonstrate that both proposed methods (DART and DRIFT) successfully reconstruct the structural information, closely matching the fully sampled ground truth data despite undersampled k-space.
 \begin{figure*}[!htb]
    \centering
    \includegraphics[width=0.85\linewidth]{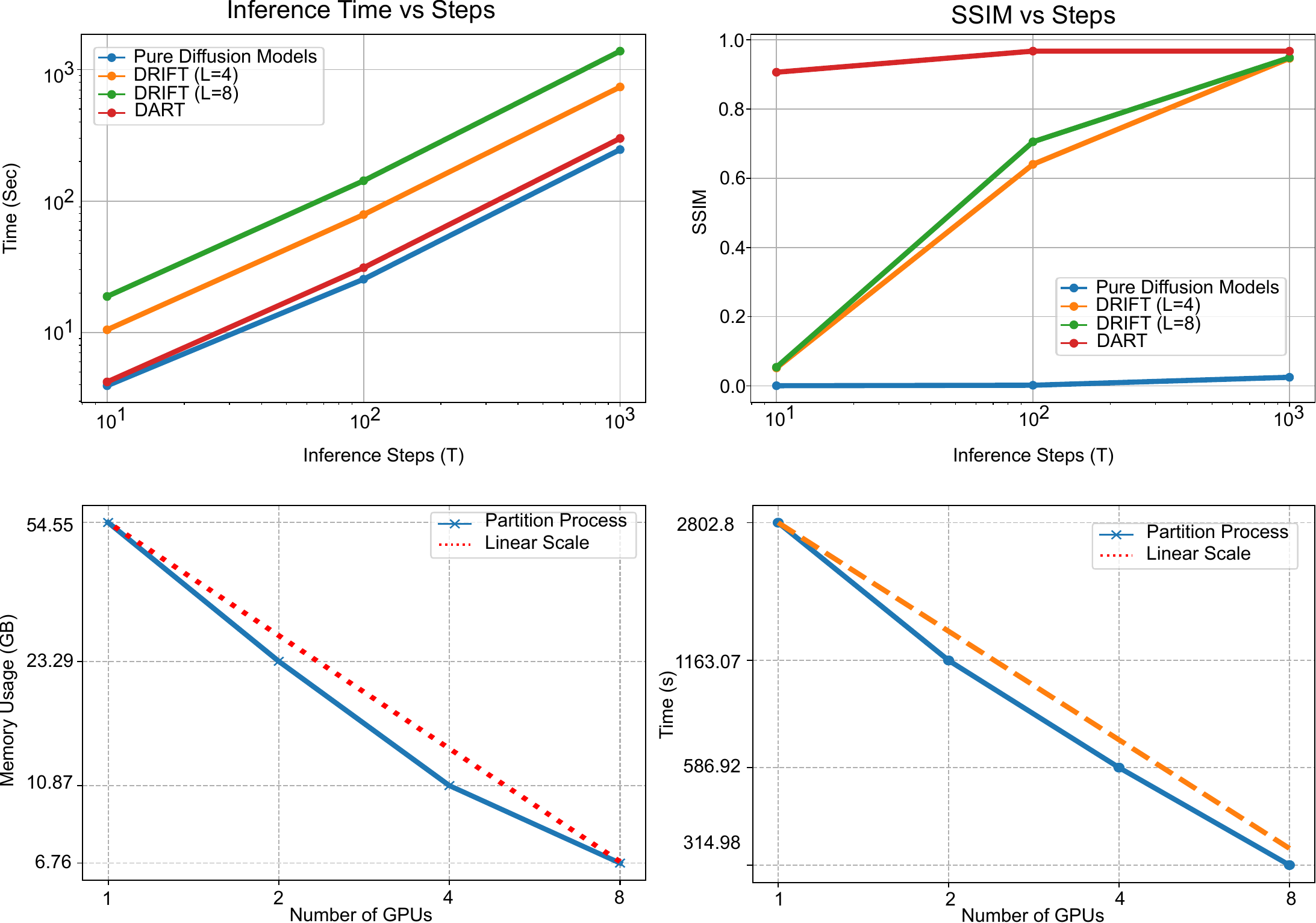}
    \caption{Top row: run time and SSIM of pure (vanilla) diffusion models \cite{ho2022video}, DART, and DRIFT. The experiments are conducted on a distributed process with 8 GPUs for multi-slice MRI data with dimension $155 \times 240 \times 240$. Bottom row: time and memory benchmarking for DART using various numbers of GPUs.}
    \label{fig:ablation_time_memory}
\end{figure*}
\AB{\begin{table*}[htb!]
    \centering
    \renewcommand{\arraystretch}{1.2}
    \begin{tabular}{|c|cc|cc|cc|}
        \hline
        \multirow{2}{*}{\textbf{\# Generated}} & \multicolumn{2}{c|}{\textbf{FVD-i3D} ($\downarrow$)} & \multicolumn{2}{c|}{\textbf{FVD-VideoMAE} ($\downarrow$)} & \multicolumn{2}{c|}{\textbf{JEDi} ($\downarrow$)} \\
        \cline{2-7}
        & \textbf{3D (single)} & \textbf{2-GPU (stack)} & \textbf{3D (single)} & \textbf{2-GPU (stack)} & \textbf{3D (single)} & \textbf{2-GPU (stack)} \\
        \hline
        50  & 2,099.01 & 2,638.97 & 251.70 & 751.61 & 7.74 & 9.64 \\
        100 & 2,109.51 & 2,675.64 & 261.79 & 762.91 & 7.49 & 9.54 \\
        150 & 2,187.89 & 2,653.14 & 263.68 & 759.70 & 7.36 & 10.06 \\
        200 & 2,051.56 & 2,653.39 & 257.39 & 753.80 & 7.76 & 9.92 \\
        250 & 2,036.96 & 2,626.47 & 256.36 & 749.66 & 7.78 & 9.81 \\
        300 & 2,018.10 & 2,584.11 & 252.91 & 750.68 & 7.71 & 9.67 \\
        366 & 2,034.72 & 2,615.10 & 254.56 & 754.38 & 7.76 & 9.79 \\
        \hline
    \end{tabular}
    \caption{FVD and JEDi for MRI multi-slice generated data from vanilla diffusion model. 3D (single) denotes single-GPU generation of \(40 \times 240 \times 240\). 2-GPU (stack) denotes distributed generation of two \(20 \times 240 \times 240\) stacks concatenated to \(40 \times 240 \times 240\). Lower is better.}
    \label{tab:FVD-JEDI}
\end{table*}}
\subsection{4D-Scanning Transmission Electron Microscopy}
\label{subsec:stem_inference} For evaluation, $30$ multi-slices ground truth crystalline materials are generated. The intensity of diffraction patterns is acquired with total scan $R = R_x \times R_y = 80 \times 80$ with detector's dimension $80 \times 80$.  Table \ref{tab:mean_ssim_mri} presents a comparison of SSIM values across other methods, namely Sparse Decomposition \cite{bangun2022inverse}, 3PIE \cite{maiden2012ptychographic}, and Torchslice \cite{diederichs2024exact}. 
The phase projection of materials is used to evaluate the SSIM. DRIFT achieves the highest mean SSIM, suggesting that it delivers the most accurate reconstruction quality by leveraging a trained diffusion model before generating promising initial guesses and refining them with a physics-based iterative solver.  One of the reasons DART performs worse in 4D-STEM case is that, unlike the MRI case, where for each slice we have a unique pair of k-space images, in 4D-STEM, the measurement data is  only the projected intensity, which leads to slow convergence.  Figure \ref{fig:BraTS_Crystal_result} presents 2D phase projections for two cubic crystalline materials, namely CoPt$_3$ and Tb$_3$InC. Each projection is annotated with its SSIM value, reflecting the structural similarity to the ground truth.  SOTA algorithms also \LV{achieve} high SSIM for this dataset but lower SSIM on average.
\subsection{Out-of-Distribution Data}
{We additionally validate our methods against baselines on out-of-distribution (OOD) data, which is critical for real-world deployment when large annotated datasets are not available. In Table \ref{tab:mean_ssim_mri_ood}, we present the evaluation results on the plant roots dataset \cite{schulz20123d} and hexagonal crystalline materials from Material Projects \cite{jain2013commentary}.}
 Both DART and DRIFT perform reconstruction using models pre-trained on MRI BraTS and cubic crystal data, with an $8\times$ acceleration factor mask based on a Gaussian distribution to undersample MRI measurement data (k-space). DART and DRIFT maintain higher SSIM under OOD settings for both MRI and 4D-STEM, respectively, indicating better generalization. Figure \ref{fig:ood_results} shows example reconstructions, where both DART and DRIFT further preserve both structural and intensity information, even when tested on out-of-distribution data.

\subsection{Ablation Studies} 
Incorporating diffusion models into scientific imaging requires consideration of the physical characteristics of the data acquisition process. 

Figure \ref{fig:ablation_time_memory} shows the trade-off between inference time and the quality of reconstruction given measurement data for vanilla diffusion model and the proposed algorithms across $8$ GPUs. In the absence of physical constraints and measurement data, the vanilla diffusion model fails to produce reconstructions that are faithful to the ground truth, as indicated by lower SSIM. It can be observed that DART increases a small fraction of the run time for inference step $ T \in \{10,100,1000 \}$. However, for DRIFT, the run time highly depends on the number of images we generate. 

We also show that distributing the data to more GPUs helps to speed up the inference process and reduce memory allocation.  The memory demand for the vanilla diffusion model with multi-slice dimension $155 \times 240 \times 240$ is very high, i.e., $54.55$ GB. This might exceed many common GPU configurations and lead to out-of-memory issues. Therefore, partitioning the job across multiple GPUs not only produces a speed up but also helps to keep per-GPU memory usage within our limits. 

 \subsection{\AB{Quantitative Evaluation of Generated Multi-Slice}}
\label{subsec:fvd}

\AB{Beyond time and memory benchmarks, we assess the distributional similarity between multi-slice images generated from a vanilla partitioned diffusion model and ground-truth using FVD (Fréchet Video Distance) \cite{unterthiner2019fvd, Ge_2024_CVPR} and JEDi (JEPA Embedding Distance) \cite{luo2024beyond}. In the absence of physics-based conditioning, a vanilla diffusion model stochastically generates multi-slice from the learned prior, so distribution-level metrics such as FVD and JEDi are the appropriate measure. We report three variants: FVD with i3D features, FVD with VideoMAE features, and JEDi with V-JEPA features. Evaluations span 50–366 generated multi-slice \LV{data} under two configurations: (i) single-GPU generation of \(40 \times 240 \times 240\) volumes; (ii) 2-GPU distributed generation that produces two independent \(20 \times 240 \times 240\) stacks and concatenates them into \(40 \times 240 \times 240\).}

\AB{Table~\ref{tab:FVD-JEDI} shows that the single-GPU configuration consistently outperforms the 2-GPU stacked setup across all sample sizes. The gap is largest for FVD with VideoMAE features, moderate for FVD-i3D, and smallest for JEDi, indicating that concatenating independently generated \(20\)-slice stacks degrades space–time coherence captured by FVD, while JEDi is comparatively robust. Although the distributed strategy performs worse, the absolute differences are limited, suggesting it remains viable under resource constraints. However, as we show in Figure \ref{fig:ablation_time_memory}, a vanilla 3D diffusion model without physics-based constraints (e.g., measurement models such as k-space in MRI or diffraction patterns in 4D-STEM) can yield non-unique reconstructions for exact ground truth data.}

\section{Conclusion}
\AB{This work introduces a physics-guided framework that integrates partitioned diffusion priors for multi-slice reconstruction in scientific imaging. In particular, we propose two algorithms—DART: Diffusion–Alternating Multi-slice Reconstruction Technique and DRIFT: Diffusion–Refined Initialization for Multi-slice Reconstruction. By coupling learned diffusion priors with explicit forward models (e.g., k-space for MRI and diffraction patterns for 4D-STEM), our framework provides a principled bridge between data-driven generative modeling and physics-based reconstruction, yielding outputs that are both perceptually plausible and physically consistent. The proposed methods are not only computationally and memory efficient, but also effective in both MRI and 4D-STEM applications, especially for general real-time multi-slice reconstruction in scientific imaging applications. Potential directions include training and evaluation on experimental 4D-STEM diffraction patterns.}

\section*{Acknowledgment}
 The authors gratefully acknowledge computing time on the supercomputer JURECA\cite{thornig2021jureca} at Forschungszentrum Jülich under grant  \texttt{delia-mp}.

\bibliography{strings}
\end{document}